# Enhancement in Photoluminescence of Pt/Ag-Pt Embedded ZrO2 Thin Films by Plasma Co-sputtering

Shailendra Kumar Mishra[1], Ibnul Farid[2], Aritra Tarafder[2,3], Joyanti Chutia[2*], Subir Biswas*[2,3], Arup Ratan Pal[2,3], Neeraj Shukla[1]

1. Department of Physics, National Institute of Technology Patna-800005, Bihar, India
2. Physical Sciences Division, Institute of Advanced Study in Science and Technology Guwahati-781035, Assam, India
3. Academy of Scientific and Innovative Research (AcSIR), Gaziabad 201002, U.P., India
* Corresponding authors' mail Id: joyanti_c@yahoo.com; subirb@iasst.gov.in

## Abstract

Platinum, Silver-Platinum embedded Zirconia (Pt/Ag-Pt ZrO2) thin films have been fabricated on silicon wafers and glass substrates using the plasma co-sputtering method. Zirconia thin films are of significant technological importance due to their remarkable electrical, optical, and mechanical properties, as well as their high melting temperature of 2715°C, which makes them increasingly attractive for various applications. In this study, ZrO2 thin films were deposited for 3 minutes, followed by the deposition of Pt-Ag/Pt onto the fabricated zirconia thin films, with deposition times ranging from 15 to 60 seconds. The varying deposition times of Pt-Ag/Pt influenced the optical and electronic properties of the thin films due to alterations in their surface roughness. The characteristics of the grown zirconia and Pt/Ag-Pt sputtered zirconia nanostructures were investigated using Atomic Force Microscopy (AFM), Scanning Electron Microscopy (SEM), X-ray Diffraction (XRD), UV-visible spectroscopy, and Photoluminescence spectroscopy. The optical transmittance of these thin films was examined across the visible and near-infrared spectral ranges. The investigation revealed various properties, such as enhanced photoluminescence and the emergence of new peaks in the visible range spectra. Plasmonic peaks were induced, and an increase in the sharpness of these peaks was observed between 403.15 nm and 512.10 nm for the Pt/Ag-Pt deposited samples. This enhancement in photoluminescence is attributed to the plasmonic properties of Pt-Ag nanoparticles on the zirconia thin film. The study demonstrates that these optically tuned thin film coatings, with their enhanced photoluminescence properties, can significantly improve the heat-resistance capacity of devices, mitigating issues related to overheating and device shutdown. Such thin films have potential applications in photovoltaic devices, optical devices, and biosensors, among others.

## Keywords

# 1. Introduction

Zirconia (ZrO2, Zirconium oxide) thin films play a significant role due to their high chemical stability, excellent corrosion resistance, and microbial resistance [1,2]. These properties, along with their electrical, optical, and mechanical characteristics, are gaining increasing attention, particularly because of zirconia's high melting temperature of 2715°C [3]. Zirconia is a wide band gap (3.10 to 5.79 eV) p-type semiconductor with abundant oxygen vacancies on its surface, making it useful as a catalyst owing to its high ion exchange capacity and redox activities [4]. ZrO2 thin films are also potential dielectric materials for applications as insulators in transistors and nano-electronic devices [5]. Zirconia exists in three well-defined crystal structures: cubic (c-ZrO2), monoclinic (m-ZrO2), and tetragonal (t-ZrO2). The monoclinic phase is stable up to 1170ºC, the tetragonal phase forms between 1170ºC and 2370ºC, and the cubic phase appears above 2370ºC. Zirconia thin films are considered potential materials for transparent electronic devices due to their high transmittance, large band gap, semiconducting behaviour, and high heat resistance.

Various methods have been used to fabricate zirconia thin films, including the sol-gel method, precipitation method [5,6], vapor phase method [4], pyrolysis method, spray pyrolysis method, hydrothermal method,[7] and sputtering method [8-10]. Zirconia applications include oxygen gas sensors, high-temperature micro-chemical sensors [11], components of solid oxide fuel cells [12], biological ceramics [2,13,14], photodetectors [15], memory devices, photoluminescence studies [16], and the enhancement of catalytic behaviour of thin films [17,18]. A visual and SPR sensor for zirconium has been developed based on Zr-induced aggregation and the change in surface plasmon resonance (SPR) absorption spectra of ATP-stabilized gold nanoparticles (Au NPs) solutions [19]. Noble metal nanostructures like platinum and silver are of significant interest due to their tunable optical and electronic properties. Platinum nanoparticles have high melting points (1770ºC), excellent corrosion resistance, high wear and tarnish resistance, and chemical stability. Silver nanoparticles (Ag NPs) are chosen for their good electrical conductivity and their ability to absorb and scatter light efficiently due to their surface plasmon frequency in the visible range [20].

Ag and Pt nanoparticles exhibit localized surface plasmon resonance (LSPR), which is crucial for applications in photodetection, photocatalysis, light harvesting, energy transfer processes, waveguiding, and chemical and biological sensing [21-23]. Plasmonic nanostructures can convert collected light into electrical energy by generating hot electron-hole pairs [24]. When excited, absorption in the nanostructure induces LSPR, leading to the production of highly energetic electrons, or hot electrons, which enhance the material's photoluminescence properties [24]. This forms a metal-semiconductor junction that improves the performance of photovoltaic and photocatalytic devices [25-29].

Currently, various thin films are used for different types of coatings to improve surface quality and functionality, especially for optical elements. These coatings protect surfaces from internal and external damage. Studies, such as those by Lee et al., have explored optical thin films for display and lighting applications. Zirconia thin films are also used as optical coatings to reduce unwanted reflected light. Flat panel displays (FPDs) carry anti-reflection (AR) coatings to avoid blurred images and reflections. Adding nanoparticles increases the diffusive property, scattering unwanted light, and enhancing clarity and resolution. Absorbing materials in the coating layers provide AR properties, albeit with a nominal decrease in transmittance [30].

In the present work, the tunable optical properties of zirconia have been explored using the sputtering technique. It was observed that the film quality, particularly photoluminescence, improved with the addition of impurities (Pt/Ag-Pt) to the zirconia thin film. The aim is to create an optically active (plasmonic) thin film with good heat-resisting capacity for optoelectronic applications. The sputtering method was chosen for its excellent uniformity, low impurity levels, highly controlled directionality, low absorption, and improved adhesion of thin films to various substrates [22,23]. Due to the deposition of Pt/Ag-Pt over the zirconia thin film, the optical properties improved. The study investigated different properties of ZrO2 thin films with silver and platinum nanoparticles sputtered over them in the presence of Ar, varying the sputtering time while keeping other deposition parameters (such as internal chamber gas pressure ratio, power supply, target, substrate, and distance between target and substrate) constant.

## 2. Materials and methods

### 2.1. Synthesis of Pt-Ag/Pt-ZrO$_2$ thin film

Thin films of Pt/Ag-Pt embedded ZrO2 (Pt/Ag-Pt ZrO2) are fabricated using the plasma co-sputtering method in a magnetron sputtering chamber with a diameter of 40 cm and a height of 30 cm. The schematic diagram of the magnetron co-sputtering chamber is shown in Fig. 1. It features three planar magnetron target holders, separated by an angle of 45°. In the present experiment, only two holders are used, designated as 'Magnetron 1' and 'Magnetron 2' in Fig. 1. The substrate is placed 5 cm below the targets. A RF power supply (2 MHz, 600 W, SEREN, Model: AT6) and a DC power supply (0-1000 V, 2 A, Hind High Vac, Model: PS-2000) are used to drive Magnetron 1 and Magnetron 2, respectively.

First, a zirconia ($ZrO_2$) thin film is deposited for 3 minutes on glass/silicon wafer substrates by placing a Zr target (diameter ~2 cm, purity 99.99%) at Magnetron 2 in an argon and oxygen plasma environment. The deposited zirconia film is designated as sample 'S1'. Pt/Ag-Pt is then sputtered onto the zirconia thin films in an Ar plasma environment to obtain Pt/Ag-Pt embedded $ZrO_2$. A Pt target (diameter ~2 cm, purity 99.99%) and an Ag target (diameter ~2 cm, purity 99.99%) are placed on Magnetron 1 and Magnetron 2, respectively. RF power of 55W is used to deposit Pt, while DC power of 65W and 70W are applied for the deposition of Ag and $ZrO_2$, respectively.

To evacuate the chamber, a diffusion pump (Hind High Vac, Model: QSV-6) along with a rotary pump (Hind High Vac, Model: 01FD-0201216) are used, achieving a base pressure of $\sim 10^{-6}$ mbar. For the deposition of the zirconia thin film, argon (Ar) (99.99%) is first introduced into the chamber via a needle valve to achieve an Ar partial pressure of $2.0 \times 10^{-4}$ mbar, followed by the introduction of oxygen ($O_2$) gas (purity 99.99%) to reach a working pressure of $1.6 \times 10^{-1}$ mbar. For the sputtering of Pt/Ag-Pt on the zirconia film, the working pressure of Ar is maintained at $1.6 \times 10^{-1}$ mbar. Pt/Ag-Pt $ZrO_2$ films are fabricated by varying the sputtering time (15s, 30s, and 60s) while keeping all other deposition parameters constant. Pt $ZrO_2$ samples for sputtering times of 15s, 30s, and 60s are designated as samples S2, S3, and S4, respectively, whereas Ag-Pt $ZrO_2$ samples are designated as S5, S6, and S7, respectively. The flow rate of the input gas is controlled using a Mass Flow Controller (Type 247D, MKS). All the deposition parameters for the various samples are summarized in Table 1.

**Table 1:** *Parameters during sample fabrication process (Base pressure~$10^{-6}$ mbar)*

| Sr. No. | Sample Name | Pressure | Deposition Time (sec) | Power (DC/RF) (W) |
|---|---|---|---|---|
| S1: $ZrO_2$ | | | | |

| 1. | S1: ZrO$_2$ | Ar: 2×10$^{-4}$ mbar<br>Ar+O$_2$: 1.6×10$^{-1}$ mbar | Zr-180sec | 70 (DC) |
|---|---|---|---|---|
| **S2-S4: ZrO$_2$/Pt** | | | | |
| 2. | S2 | Ar: 1.6×10$^{-1}$ mbar | Zr-180sec<br>Pt-15sec | 70 (DC)*<br>55 (RF)** |
| 3. | S3 | | Zr-180sec<br>Pt-30sec | |
| 4. | S4 | | Zr-180sec<br>Pt-60sec | |
| **S5-S7: ZrO$_2$/ Ag-Pt** | | | | |
| 5. | S5 | Ar: 1.6×10$^{-1}$ mbar | Zr-180sec<br>Ag/Pt-15sec | Zr-70 (DC)*<br><br>Ag-65 (DC)*<br>Pt-55 (RF)** |
| 6. | S6 | | Zr-180sec<br>Ag/Pt-30sec | |
| 7. | S7 | | Zr-180sec<br>Ag/Pt-60sec | |
| (DC)*: DC Power supply (RF)**: RF | | | | |

## 2.2. Characterization techniques

X-ray diffraction (XRD) analysis (D8 Advance, Bruker AXS, with Cu Kα radiation, λ = 1.5406 Å) has been performed on the deposited samples to determine their structural properties. To investigate the surface morphology of the fabricated films, Atomic Force Microscopy (AFM) (NTEGRA Prima, NT-MDT instrument, using a silicon cantilever in semi-contact mode) analysis has been carried out. Surface morphology, particle size, and elemental stoichiometric ratios of the zirconia films and the Pt/Ag-Pt sputtered zirconia thin films are determined using Scanning Electron Microscopy (FESEM, SIGMA VP, ZEISS) and Energy Dispersive Spectroscopy (EDS), respectively.

The optical properties (absorbance and bandgap) of the deposited films are examined using a UV-Vis Spectrophotometer (UV-2600, Shimadzu, Japan). Photoluminescence (PL) analysis is conducted at three different excitation wavelengths (290 nm, 360 nm, and 380 nm) to assess changes in optical absorption and emission using a PL spectrometer (HORIBA CANADA QM-8075-21-C).

## 3. Results

### 3.1. X-ray Diffraction (XRD) - Analysis

*(a). Sample Pt-ZrO$_2$*

**Figure 1**: Panels i and ii(a-c) exhibit the X-ray diffraction (XRD) analysis of ZrO2 thin films and Pt-sputtered ZrO2 thin films at different deposition times. From Figure 1(i), it is evident that the zirconia films possess both tetragonal (t) and monoclinic (m) structures, with monoclinic structures being more dominant. The transition crystal structures and the integrated intensities of various planes, such as m(110) at a 2θ value of 23.86°, 26.05° m(111), 27.44° m(110), 30.18° t(101), 32.63° m(111), 37.68° t(201), 39.26° m(111), and 39.80°, m(112) at 41.03°, m(211) at 43.76°, and m(112) at 50.56°, have been observed [24,25]. Monoclinic structures exhibit intense sharp peaks at 26.05°, 27.63°, 32.63°, and 39.26° with a smaller peak at 50.56°. Numerous small tetragonal peaks are present within the 2θ range of 23.86°-50.20° [6].

For sample S2, as shown in Figures 1(ii,a) and 1(ii,b), peaks corresponding to both tetragonal (t) and monoclinic (m) structures are observed within the 2θ range of 23.86°-50.98°. When platinum is sputter deposited over the zirconia film for 15 seconds and 30 seconds, peaks at 26.05° and 39.67° become sharper, indicative of an increased monoclinic structure orientation of Pt and a random, nonuniform arrangement of zirconia and platinum. The sharp monoclinic peak of zirconia diminishes at 27.31° and shifts towards 39.03°, attributed to lattice contraction due to the addition of platinum.

In Figure 1(ii,c), under similar deposition parameters, zirconia thin films are deposited, followed by Pt sputter deposition for 60 seconds. There appears to be no significant change in the crystal structures and peaks of the deposited sample. Platinum peaks become sharper with increased deposition time, with a prominent peak at 39.85°, attributed to a higher concentration of platinum atoms over the zirconia thin film [26,36].

*(b). Sample Ag/Pt-ZrO$_2$*

**Figure 1**: Panels iii(a-c) present the X-ray diffraction (XRD) modes of Ag-Pt co-sputtered ZrO2 thin films at different deposition times. The transition crystal structures formed by Ag-Pt co-sputtering on ZrO2 thin films show integrated intensities at the following 2θ values: m(110) at 25.07° and 34.51°, m(111) at 32.69°, 41.72°, and 69.10°, m(112) at 45.30° and 45.40°, t(110) at 29.23° and 64.61°, and t(200) at 73.83°. Monoclinic structures exhibit sharp peaks at 25.16°, 30.39°, 32.51°, and 38.56°, while tetragonal structures show peaks at 38.74°, 65.21°, and 73.72° [6].

Panels iii(a-c) in Figure 1 provide information about samples S5-S7. Sharp Ag peaks are formed, and peak shifts are observed between 45.20°-45.47° and 69.12°-72.84° due to agglomerated Ag/Pt nanostructures. The deposition is non-uniform over the zirconia thin film, resulting in several peaks from the deposited elements (Zr, ZrO2, Pt, and Ag). The grain size (D) of the ZrO2 and Ag-Pt ZrO2 thin films has been estimated using Debye–Scherrer's formula. The results indicate broad diffraction peaks, signifying that the grain size of the fabricated thin film grains is in the nanometre range.

The grain size (D) of the ZrO2 and ZrO2-Pt thin films has been estimated using Debye–Scherrer's formula.

$$D = \frac{k\lambda}{\beta \cos\theta} \quad (1)$$

Where λ is the X-ray wavelength, β is the full width at half maximum (FWHM) of XRD peaks, and θ is the diffraction angle. The grain size (D) of the samples in Figure 1: i(a, b, c, d) is estimated to be 9.11 nm, 9.18 nm, 9.30 nm, and 9.80 nm, respectively. For the samples S5, S6, and S7 in Figure 1: iii(a, b, c), the grain sizes are found to be 10.80 nm, 11.30 nm, and 11.92 nm, respectively [36,37].

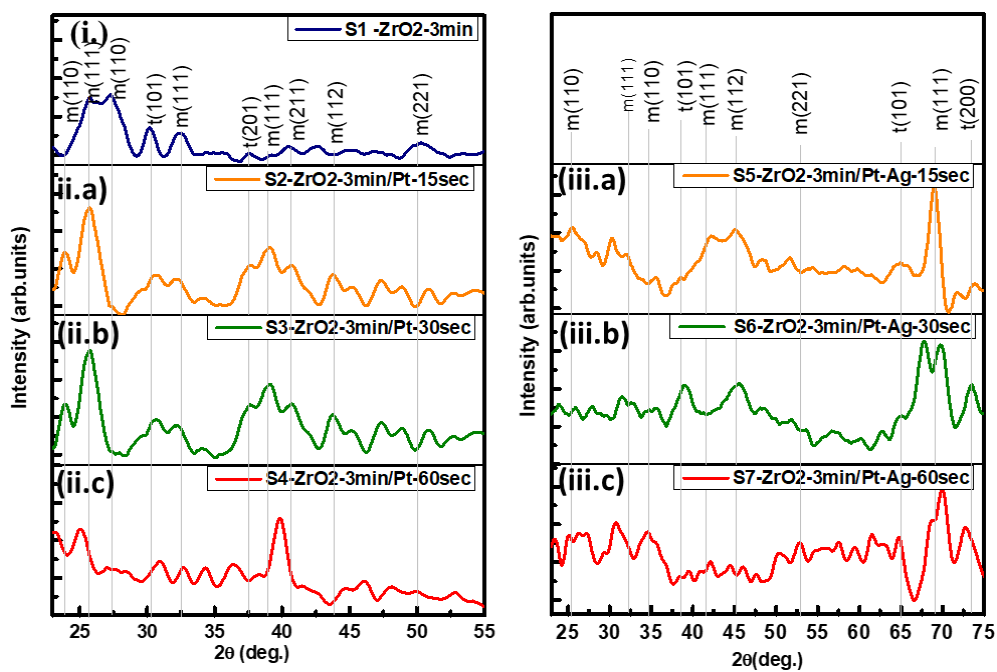

**Fig.1(i, ii, iii):** *XRD graphs of i zirconia ii (a-c) Pt-NPs sputtered zirconia thin films and iii (a-c) Pt/Ag-NPs sputtered zirconia thin films at various time intervals.*

### 3.2. Atomic Force Microscopy (AFM) Analysis

To investigate the surface morphology of the fabricated films, systematic AFM analysis has been carried out on samples prepared in various batches.

#### 3.2.1. AFM analysis of *Pt-ZrO$_2$ samples*

The AFM micrographs exhibit the formation of conical nanopillar structures in the thin film, each with different height profiles (AFM micrographs attached to supplementary data) [38]. A gradual but systematic decrease in the roughness of samples (S1, S2, S3, and S4) is observed with an increase in deposition time. From the analysis, the heights of the nanopillars in samples i, ii(a), ii(b), and ii(c) are observed to be 10 nm, 20 nm, 25 nm, and 40 nm, respectively. The thickness of the film increases due to the continuous nucleation and growth of the particles. Thus, AFM micrograph analysis suggests that the thin films follow the Stranski-Krastanov

(SK) growth mode, also known as "Layer-Plus-Island Growth."

The root mean square (RMS) values of roughness, estimated from Nova software with a resolution of 1 µm, are 1.16 nm, 5.82 nm, 0.60 nm, and 0.43 nm for samples i, ii(a), ii(b), and ii(c), respectively. The number of sharp conical pillars and the surface area of the thin film decrease with increasing deposition time. Continuous deposition of Pt-NPs results in an increase in the diameter of the conical pillars, rendering the pillars indistinct.

### 3.2.2. AFM analysis of *Ag/Pt-ZrO₂ samples*

The AFM micrographs in Figure 2: iii(a), iii(b), and iii(c) demonstrate the formation of conical nanopillar structures in the Ag-Pt NPs sputtered Zirconia thin film samples, with varying deposition times of Ag-Pt NPs. The shape of the pillars and the roughness differ across the different deposition times. The heights of the pillars are estimated to be 30 nm, 60 nm, and 200 nm for samples iii(a), iii(b), and iii(c), respectively. The RMS values of roughness are found to be 1.16 nm, 3.41 nm, 9.10 nm, and 26.49 nm. Uniform and sharp conical pillars are formed with increased deposition time, resulting in an increase in the surface area and roughness of the film.

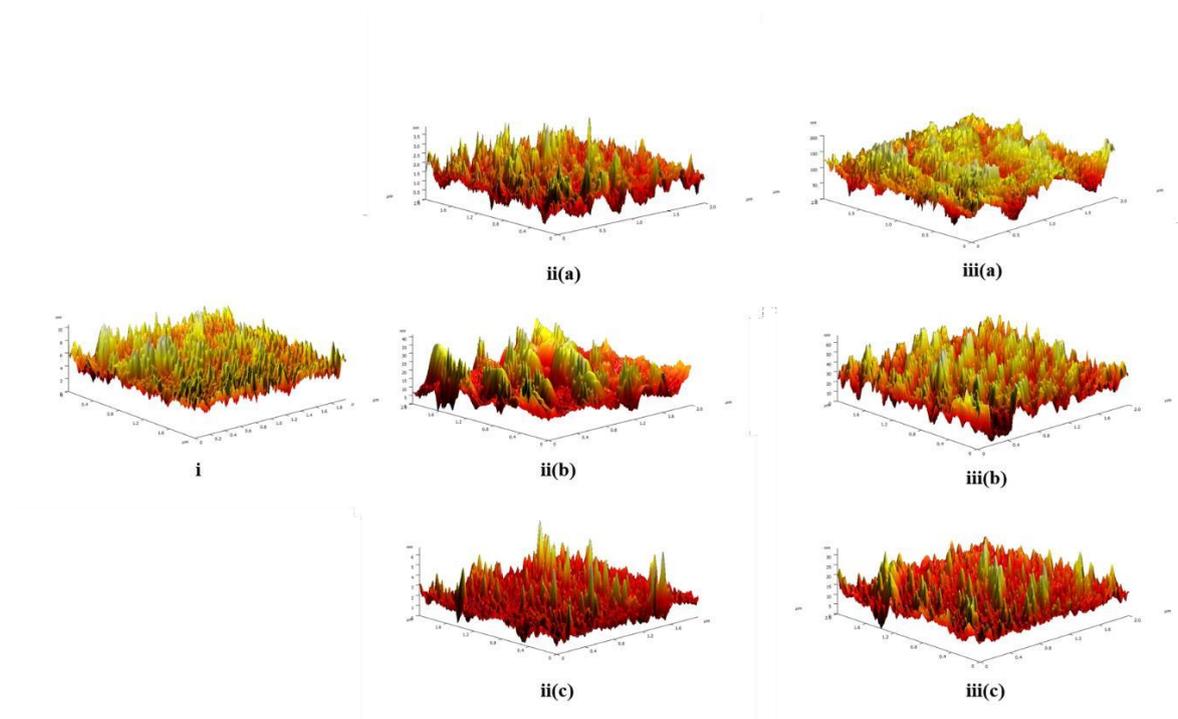

**Fig.2: (i, ii, iii):** *AFM micrographs of i-zirconia, ii (a-c) Pt-NPs sputtered zirconia thin films and iii (a-c) Pt/Ag-NPs sputtered zirconia thin films at various time intervals.*

It The thickness of the film increases with longer deposition times, as observed from the analysis above. This is attributed to increased collisions between the inserted Ar gas in the chamber and sputtered atoms, reducing the mean free path and thereby lowering the kinetic energy of ejected atoms. Consequently, sputtered nanoparticles (NPs) nucleate and grow, forming conical pillar-like nanostructures on the substrate as these nucleated particles deposit layer by layer.

In samples ii(a), ii(b), and ii(c), where Pt NPs are sputtered over ZrO2 thin film in the presence of Ar gas, uniform and sharp conical pillars are suppressed due to continuous particle deposition, resulting in decreased roughness. In contrast, when both Ag-Pt NPs are co-sputtered in the presence of Ar, the number of sharp conical pillars increases along with film thickness, leading to increase in the roughness of the surface of the thin film [38,39]. This behaviour may stem from slower nucleation of sputtered Pt-Ag NPs, as both metals compete for nucleation sites, enhancing overall film growth. Increased collisions within the chamber contribute to higher film roughness.

The slower nucleation of Pt-NPs and their higher ductility and larger radii compared to Ag-NPs result in easier embedding below the film surface, contributing electrons to the ZrO2 crystals. This lattice contraction due to Pt-NPs is evident in SEM images, whereas Ag-NPs are clearly visible on the film surface in SEM images below.

### 3.3. Scanning Electron Microscope (SEM) Analysis

The surface morphology, particle size, and elemental stoichiometric ratios of both Zirconia films and Pt/Ag-Pt sputtered Zirconia thin films have been characterized using SEM and EDS analyses, respectively. It has been observed that the nanostructures exhibit nearly homogeneous morphology with consistent granular grain size. The size of these particles plays a crucial role in tuning optical and electronic properties, which are responsible for phenomena like Localized Surface Plasmon Resonance (LSPR) [40], as explained below.

### 3.3.1. SEM & EDS spectra for *Pt-ZrO$_2$,* samples

Figure 3, panels i, ii, and iii(a-c), illustrates SEM images of samples S1 to S4, respectively. SEM analysis of sample S1 (Figure 3: i) shows a film-like structure with a uniform surface due to particle deposition. In samples S2, S3, and S4 (Figure 3: ii(a-c)), where Pt is sputtered over the zirconia thin film for 15s, 30s, and 60s, respectively, the zirconia particles interact and bond with Pt-NPs, resulting in systematic lattice contraction.
Specifically, in sample S2 (Figure 3: ii(b)), the initially uniform zirconia thin film starts forming a smooth, granular structure. This granular structure becomes more pronounced with increased deposition time of Pt-NPs, evident in samples S3 (Figure 3: ii(c)) and S4 (Figure 3: ii(d)). Notably, nanopore-like gaps can be observed between the granules in these samples, which are attributed to strong interactions between zirconia crystals and Pt-NPs, causing rearrangement of the zirconia granules and the formation of separated nanopore-like gaps.

### 3.3.2. SEM & EDS spectra for *Ag/Pt-ZrO$_2$ sample*s

Figure 3: iii(a-c), SEM analysis reveals the non-uniform distribution of Ag-Pt NPs co-sputtered over the zirconia thin film, forming distinct barrel-shaped crystal structures in samples iii(a-c). As the deposition time of Ag-Pt increases, there is a proportional increase in the deposition of Ag-Pt crystals, resulting in thicker and rougher thin films with a higher density of closely packed particles.

The presence of Pt in ZrO2-Pt thin films contributes to the formation of granular crystal structures, as observed. Conversely, Ag-NPs are clearly discernible on the film surfaces.

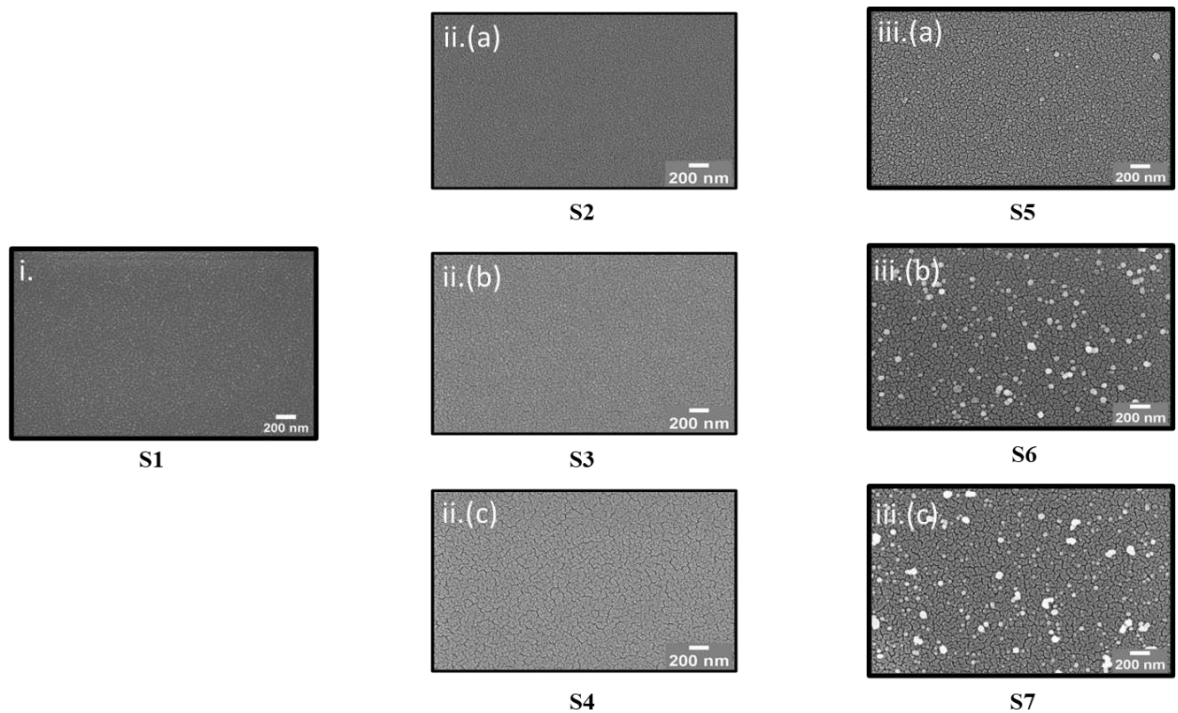

**Fig.3: (i, ii, iii):** *The SEM analysis of the samples of i-zirconia ii(a-c) Pt-NPs sputtered zirconia thin films and iii(a-c) Pt/Ag-NPs sputtered zirconia thin films at various time intervals (a) 15s, (b) 30s and (c) 60s, respectively.*

The elemental quantification of sputtered elements (Zr, O2, and Pt) was conducted using Energy Dispersive Spectroscopy (EDS) for samples S1-S7 (refer to Figure in supplementary data). These figures illustrate the presence of deposited elements on the substrate. The optical and electronic properties of the thin films depend significantly on the weight percentage and atomic percentage of these deposited elements.

During the fabrication of zirconia (ZrO2) thin films via sputtering, the ratio of zirconium to oxygen atoms can vary due to the scattering of Zr atoms in the Ar/O2 gas atmosphere. This variation affects the nucleation ratio of Zr and O2, influencing the elemental composition in the sputtered films. However, these ratios show minimal variation due to slight changes in the deposition conditions.

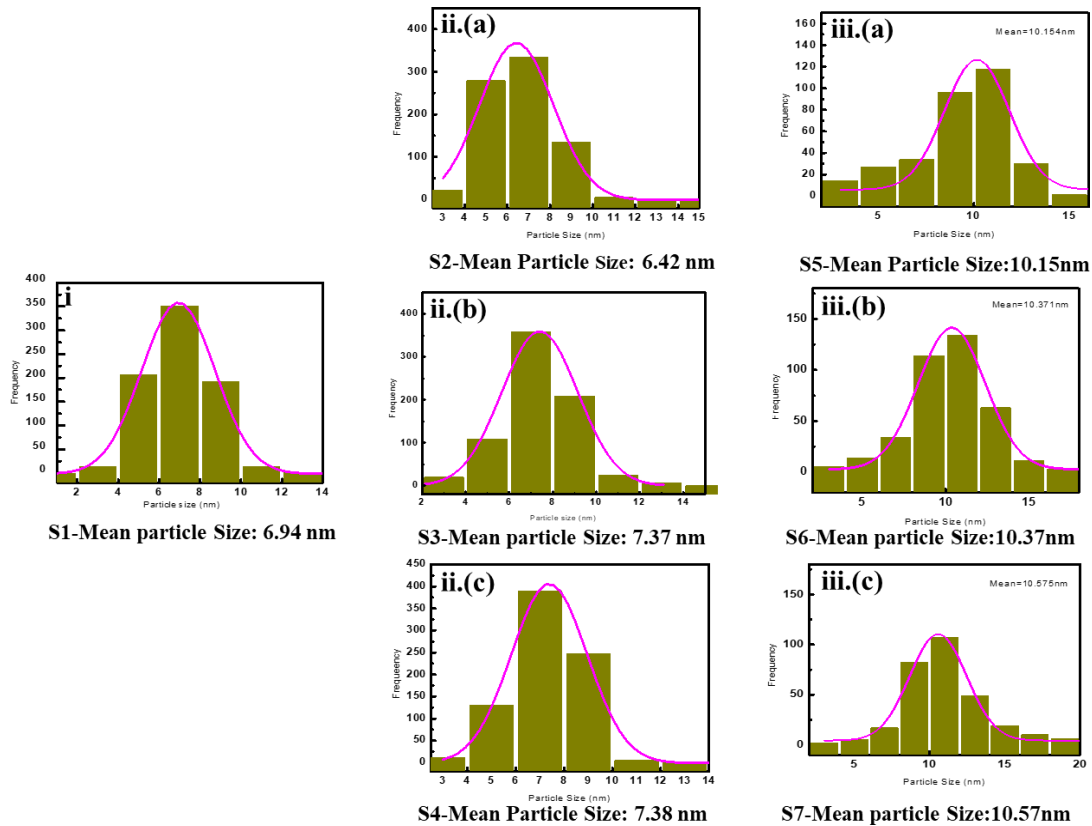

**Fig.4: (i, ii, iii):** *The particle size of the samples of (i) zirconia (ii) Pt-NPs sputtered zirconia thin films and (iii) Pt/Ag-NPs sputtered zirconia thin films at various time intervals (a) 15s, (b) 30s and (c) 60s, respectively.*

The optical and electronic properties of thin films depend significantly on particle size, which was analyzed using Image-J software. Figure 4 (i, ii, iii) illustrates the particle size distribution for samples S1, S2, S3, S4, S5, S6, and S7, respectively. Mean particle sizes were calculated using Gaussian fitting.

For sample S1, the particle size of zirconia is 6.94±0.04 nm, while Pt particles are in the range of 2-3 nm [41]. In samples ii(a), ii(b), and ii(c), where Pt is sputtered over the zirconia thin film, particle sizes are 6.42±0.06 nm, 7.37±0.02 nm, and 7.38±0.03 nm, respectively. This indicates that the average particle size follows the order S2 < S3 ≤ S4, with variations attributed to increased Pt-NPs deposition time.

For samples iii(a), iii(b), and iii(c), where Ag-Pt NPs were co-sputtered onto the zirconia thin film, particle sizes are 10.15±0.17 nm, 10.37±0.05 nm, and 10.57±0.09 nm, respectively. These sizes show a slight increase due to longer deposition times of Ag-Pt NPs. SEM images in Figure 3 (i, ii(a-c), iii(a-c)) clearly show the presence of large nucleated Ag-Pt NPs on the thin film, correlating with increased deposition times.

### 3.4. UV-Vis Spectroscopy Analysis

UV-Vis spectroscopy analysis was performed to determine the absorbance and optical band gap ($E_g$) of the deposited thin film samples. The band gap energy ($E_g$) was calculated using the Tauc relation [42], which correlates the absorption coefficient ($\alpha$) with the incident photon energy ($h\nu$). According to the Tauc relation, plotting $(\alpha h\nu)^2$ against photon energy ($h\nu$) yields a straight line for direct transition band gaps. The optical band gap energy ($E_g$) is determined by extrapolating this linear region to the energy axis and identifying the intercept [36].
This method allows for the assessment of the electronic band structure of thin film materials based on their absorption spectra, providing insights into their optical and electronic properties.

**(a) Absorption spectra and Bandgap -** *$Pt-ZrO_2$, $Ag/Pt-ZrO_2$ samples*

The Tauc plots obtained from samples S1, S2, S3, and S4 are depicted in Fig.5. For samples S1, the Tauc plot is shown in panels i(a-c), with the corresponding UV-Vis absorption spectra included in the inset graph. The band gap energy ($E_g$) shows minimal variation as the deposition time of Pt increases from 15s to 30s (samples S2 and S3), remaining within the range of 2.01 eV to 2.05 eV. However, a further increase in Pt deposition time from 30s to 60s (samples S3 and S4) results in a notable increase in the band gap from 2.05 eV to 2.67 eV.

In the absorption spectra, a redshift is observed due to electron transitions influenced by the interaction of platinum with zirconia [34]. This shift indicates a departure from the intrinsic properties of the zirconia thin film, possibly due to doping effects or changes in carrier concentration. Such optical changes reflect modifications in the electronic structure of the thin film induced by varying Pt deposition times.

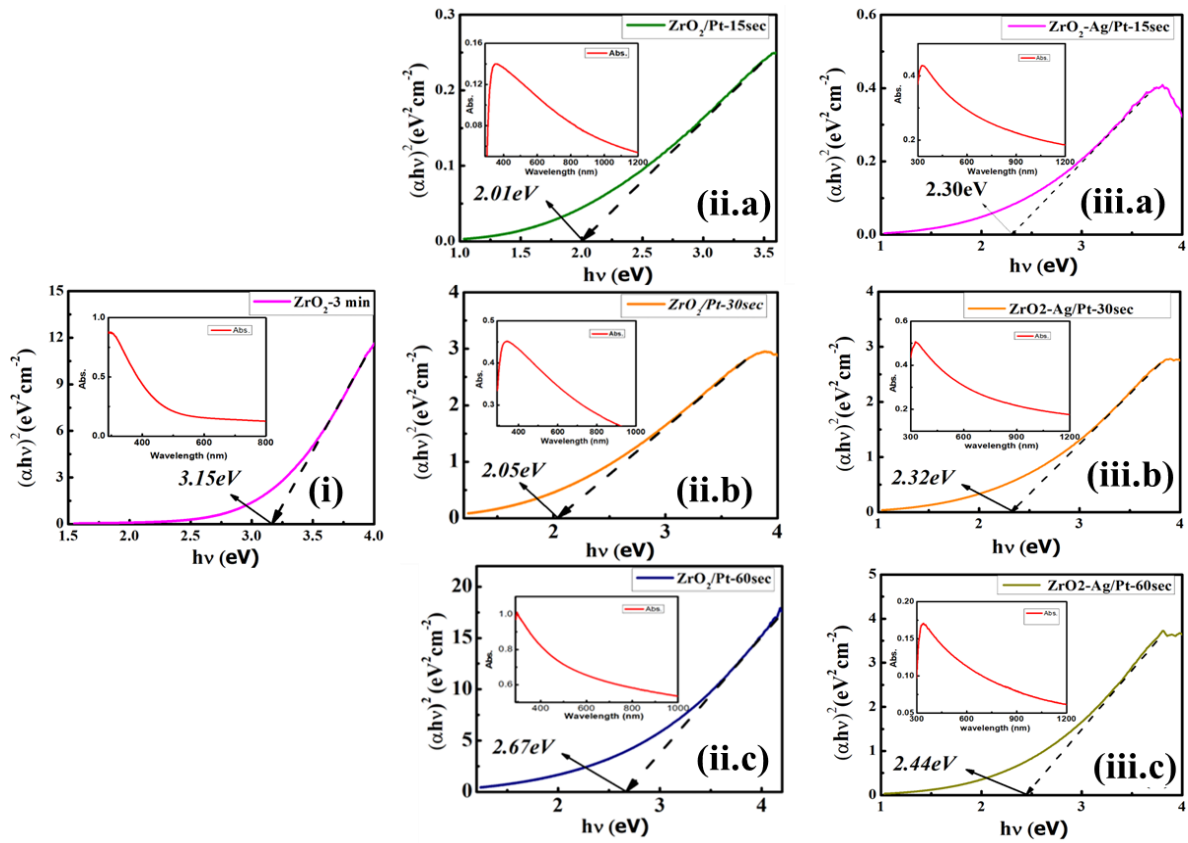

**Fig.5: (i, ii, iii):** *The Tauc plot obtained from (S1) zirconia (i) Pt-NPs sputtered zirconia thin films and (ii) Pt/Ag-NPs sputtered zirconia thin films at various time intervals (a) 15s, (b) 30s and (c) 60s, respectively and corresponding UV-Vis absorption spectra are shown in the inset graph.*

### 3.4.1. Absorption spectra and Bandgap estimation for Ag/Pt-ZrO$_2$ samples

The band gap energies deduced from samples S1, S5, S6, and S7 are 3.15 eV, 2.30 eV, 2.32 eV, and 2.44 eV respectively, as shown in Fig.5: ii(a-c). In Fig.5 ii(a) and ii(b), a decrease in band gap is observed after adding Ag/Pt-NPs over zirconia in sample S1. The band gap energy remains nearly constant as the deposition time increases from 15s to 30s, and a slight increase is noted with further deposition time from 30s to 60s.

In both categories of depositions described in sections 3.4.1 and 3.4.2, the band gap of zirconia initially decreases when Pt/Ag-Pt NPs are sputtered onto zirconia thin films for shorter durations, and then increases with longer deposition times of Pt/Ag-Pt NPs. This change in band gap can be attributed to variations in the concentration of impurities within the sample. The Burstein-Moss (B-M) effect [43] explains such behavior, where the optical band gap of degenerately doped semiconductors increases when all states near the conduction band are populated, causing the absorption edge to shift to higher energies. This phenomenon allows for tuning the optical properties of the material [44].

**Photoluminescence analysis**

Photo-luminescence (PL) analysis was conducted to investigate the optical absorption and emission changes in zirconia thin films upon the addition of impurities, specifically plasmonic materials such as Pt/Ag-Pt. Zirconia thin films were chosen as the base material for studying these changes, given their wide band gap. PL analysis was performed using excitation wavelengths of 290 nm, 360 nm, and 380 nm.

**Emission peak analysis of *Pt-ZrO$_2$, Ag/Pt-ZrO$_2$ samples***

Fig. 6: i(a) presents the photo-luminescence spectra of samples S1, S2, S3, and S4 at an excitation wavelength of 290 nm. Sharp emission peaks at 343.29 nm indicate band gap emissions in both pure ZrO2 and Pt-embedded ZrO2 nanostructures. A peak at 331.96 nm suggests the presence of sub-energy levels near the conduction band, possibly due to oxygen vacancies (Vo) in the material. These singly ionized oxygen vacancies contribute to UV emissions through radiative recombination, observed across all deposited samples [28, 34].

In samples S3 and S4, there is no significant change in band gap emission intensity at this excitation wavelength. The emission near the band gap decreases with increasing Pt-NPs over zirconia and diminishes in S2. A new peak emerges at 404.32 nm in S2, attributed to electron-phonon (e-Ph) interactions. The shorter deposition time of Pt-NPs on zirconia enhances e-Ph interactions, suppressing emission from singly ionized oxygen vacancies. Conversely, longer deposition times intensify induced emission peaks.

Embedded Pt nanoparticles may act as dopants on the zirconia thin film, facilitating electron transfer to zirconia. This interaction forms covalent bonds between vacant oxygen sites on zirconia and Pt-NPs.

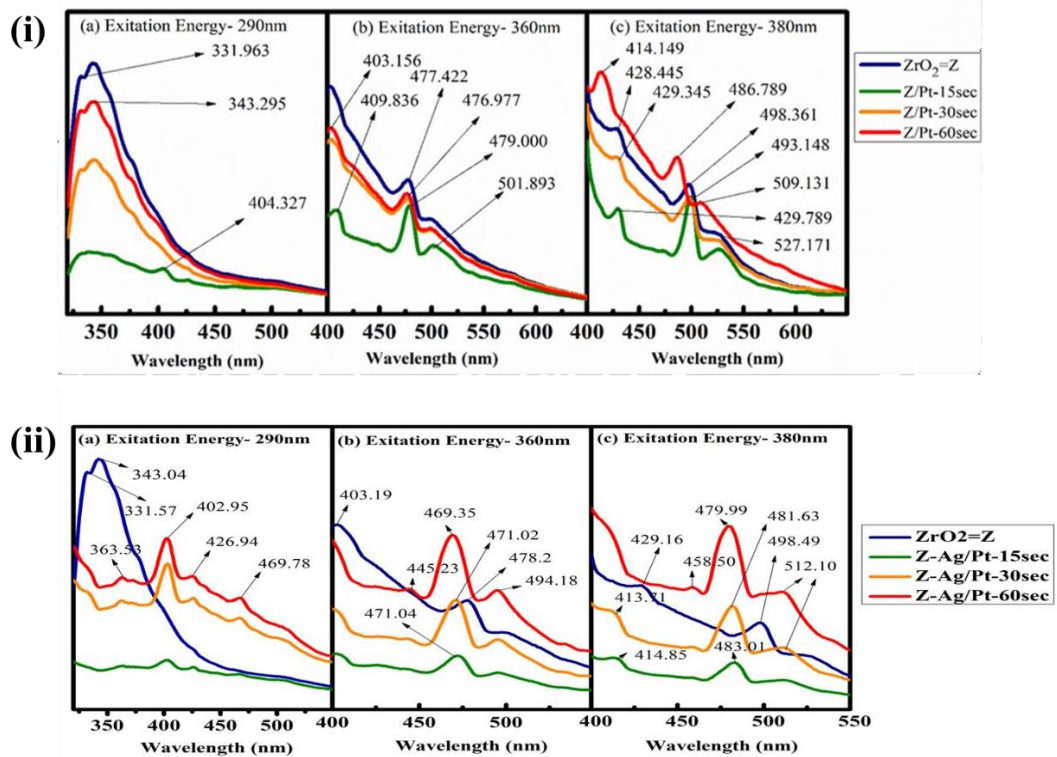

**Fig.6: (i, ii):** *PL emission peaks of (i) Zirconia, Pt-NPs sputtered zirconia thin films and (ii) Zirconia, Pt/Ag-NPs sputtered zirconia thin films for various deposition times at excitation energy (a) 290 nm, (b) 360 nm and (c) 380 nm, respectively.*

Fig. 6: i(b) displays the photo-luminescence spectra for samples S1, S2, S3, and S4 at an excitation wavelength of 360 nm. Band gap emission peaks are observed at 403.15 nm and 409.15 nm. The peak observed between 476.99 nm and 479.00 nm is attributed to single ionized oxygen vacancies in the zirconia nanostructures. In samples S1, S3, and S4, there is a slight reduction in emission intensity with no significant change in peak position. However, in sample S2, where Pt coverage on the film is lower, the PL peak sharpens at 479.00 nm and 501.89 nm due to electron-phonon (e-Ph) interactions [46]. This enhancement is attributed to stronger covalent bonding between Pt nanoparticles and the zirconia crystal, facilitated by lower Pt deposition, increasing e-Ph interactions. A gradual redshift is observed with decreasing Pt deposition time.

Fig. 6: i(c) presents the photo-luminescence spectra for samples S1, S2, S3, and S4 at an excitation wavelength of 380 nm. Band gap emission peaks are observed between 414.14 nm and 429.78 nm. The peak between 486.78 nm and 493.14 nm arises from single ionized oxygen vacancies, while shifts between these peaks are attributed to vibrational excitation of free electrons in Pt-NPs. The peak intensity sharpens with lower Pt deposition time on the film, and peaks observed between 509.13 nm and 527.17 nm indicate e-Ph coupling [46], illustrated in Fig. 14 (c) for sample S2. Increased Pt deposition rates result in a significant blue shift.

**Emission peak analysis of Ag/Pt-ZrO$_2$ samples**

**Fig. 6: ii(a)** illustrates the photo-luminescence (PL) spectra at an excitation wavelength of 290 nm for samples S1, S5, S6, and S7. In sample S1, shown in Fig. 15(a), intense sharp peaks are observed at 331.57 nm and 343.03 nm, attributed to band gap emission. However, near-band gap emission in samples S5, S6, and S7 appears less sharp or vanishes, potentially due to high excitation energy over the thin film. The presence of sub-band gaps is indicated by the division of peaks. Samples S5, S6, and S7 exhibit new sharp peaks: 402.95 nm due to single ionized oxygen vacancies (Vo) in the deposited ZrO2 nanostructures enhanced by free electrons from Pt-Ag NPs, 426.93 nm possibly due to e-Ph interaction, and 469.78 nm as a surface plasmonic peak attributed to Ag-NPs [46-48]. The sharpness of these peaks increases with longer Ag-Pt deposition times over the zirconia thin film.

**Fig. 6: ii(b)** shows the photo-luminescence spectra at an excitation wavelength of 360 nm for samples S1, S5, S6, and S7. In S1, near-band gap emission peaks are observed at 403.183 nm and a sharp peak at 478.200 nm due to oxygen vacancies (Vo) near the conduction band. Samples S5, S6, and S7 exhibit shifted peaks between 469.35 nm and 471.02 nm, sharpening with increased Ag-Pt deposition time. Peaks induced at 445.23 nm and enhanced sharpness at 469.35 nm are attributed to Ag-NPs, which exhibit plasmonic peaks. A peak at 494.17 nm is possibly due to free electrons on the film surface from Pt/Ag-NPs [47,48].

**Fig. 6: ii(c)** depicts the photo-luminescence spectra at an excitation wavelength of 380 nm for samples S1, S5, S6, and S7. In sample S1, near-band gap emission peaks are found at 429.168 nm and a sharp peak at 498.499 nm, likely due to oxygen vacancies (Vo) near the conduction band. Samples S5, S6, and S7 show shifted near-band gap peaks between 413.71 nm and 429.16 nm attributed to the Burstein-Moss effect. The onset of plasmonic peaks and enhanced sharpness at 458.50 nm and 479.99 nm are due to Ag-NPs [47,48]. An intensity enhancement at 512.10 nm is observed with increased deposition time, possibly due to free electrons induced by Ag-Pt NPs or e-Ph coupling.

The enhancement in peak sharpness and luminescence is attributed to the plasmonic behavior of Ag-NPs over the zirconia thin film. These surface plasmonic peaks may also be induced by electron-phonon interactions. PL peaks between 331.85 nm and 378.12 nm are attributed to ionized oxygen vacancies in the conduction band, where UV emission occurs from radiative recombination of a photogenerated hole with an electron occupying an oxygen vacancy (NBE transition) [45].

Transition metal oxides typically exhibit two types of emission: UV emission (Near Band

Edge; NBE-Emission) and visible emission (Deep Level Defect/Traps, DL-Emission). PL emission peaks between 100-400 nm in the UV region (blue-shift) are intense and known as the Soret (B-band), indicating electronic transitions from π-π* [49]. The splitting of peaks at an excitation wavelength of 290 nm suggests two absorption bands in the UV region, indicative of NBE-emission transitions attributed to the high crystal quality of the deposited samples.

Peaks between 400-700 nm in the visible region (red-shift) are attributed to single ionized oxygen vacancy (Vo) defects in the deposited ZrO2 nanostructures. Absorption bands in the visible region between 2.5-3 eV are known as Q-Bands and also assigned to π-π* transitions due to vibrational excitation of free electrons [49]. The induction and enhancement of peaks observed at different excitation wavelengths between 290 nm and 380 nm are due to Pt/Ag nanoparticles over zirconia, potentially facilitating Localized Surface Plasmon Resonance (LSPR) due to the smaller size of noble metal [40,50]. The observed blue shift with increasing deposition time of Ag-Pt NPs is due to their smaller size compared to the incident wavelength of light [40-50]. Co-sputtered Ag-Pt NPs on zirconia thin films exhibit better emission peaks compared to Pt-sputtered zirconia thin films [21-23,47,48].

**Proposed structure and surface analysis of the thin film**

The interaction of Pt atoms with the Zirconia (ZrO2) crystal involves the formation of partial covalent bonding. X-ray diffraction (XRD) analysis indicates that the predominant crystal structures of zirconia in samples S2, S3, and S4 are monoclinic. Zirconia is a transition metal dioxygen complex where oxygen is typically bound to zirconium in either monoclinic or tetragonal structures, integral to its lattice.

Platinum (Pt) atoms, with their larger atomic structure compared to silver, readily provide electrons that can form covalent bonds with free oxygen radicals within zirconia. In deposited ZrO2 thin films, this interaction facilitates the formation of singly ionized oxygen vacancy (Vo) defects, which can also bond covalently with Pt. SEM analysis confirms that Pt nanoparticles contribute to lattice contraction, resulting in the formation of nanopore separations within the structures and influencing lattice vibrations [46].

Similarly, silver nanoparticles (Ag-NPs) also exhibit the ability to donate free electrons. Their strong interaction with light arises from their surface plasmon frequency, which falls within the visible range of the electromagnetic spectrum. This phenomenon, known as Localized Surface Plasmon Resonance (LSPR), enhances photoluminescence by amplifying light absorption and emission processes [47,51].

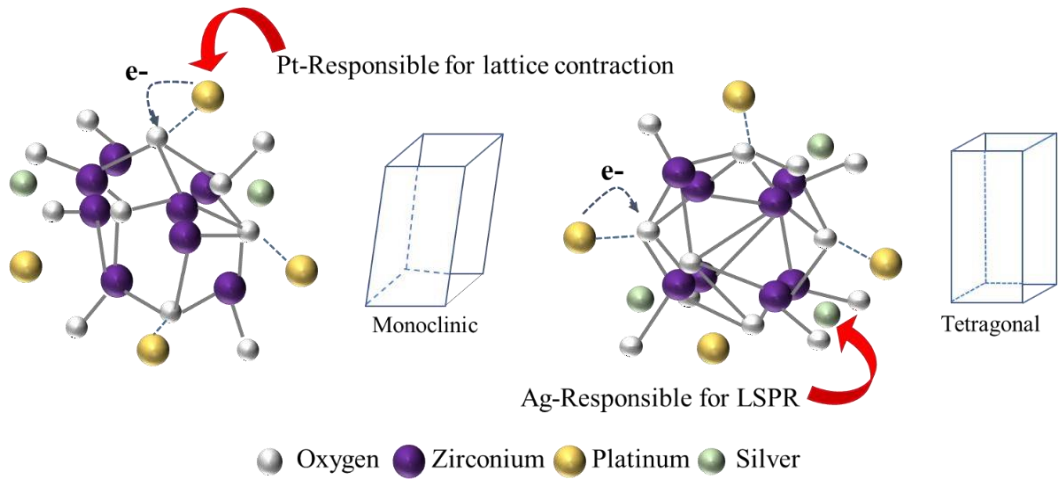

**Fig.7:** *Proposed structure of sputtered Pt-Zirconia thin film. (Zirconia structure)* [52].

Similarly, in samples S5, S6, and S7, where both Ag-Pt nanoparticles (NPs) are co-sputtered, platinum's larger atomic size facilitates easier electron donation compared to silver. However, silver also contributes to bonding with zirconia, albeit at a slower rate than platinum. SEM analysis reveals that the granular structure undergoes changes with increased Pt deposition time, with silver atoms visibly aggregating on films S5, S6, and S7.

The combined influence of Ag-Pt enhances the photoluminescence of the film, indicating that Ag-Pt embedded ZrO2 thin films exhibit surface plasmon resonance activity [50-51].

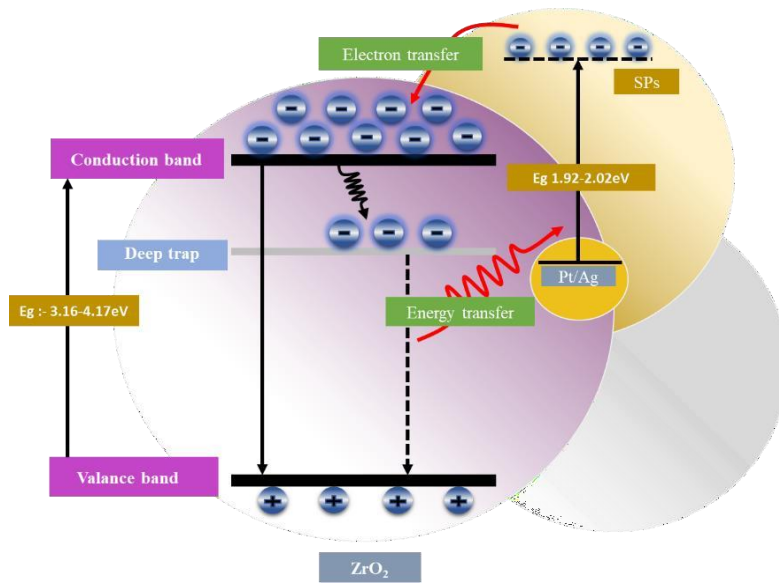

**Fig.8:** *Illustration shows the energy and electron transfer in the sputtered (Pt/Ag-Pt ZrO$_2$) thin film.*

Figure 8 illustrates the mechanism of interaction between the zirconia thin film and sputtered plasmonic materials. Oxygen vacancies (Vo) near the conduction band of zirconia trap Pt/Ag nanoparticles (NPs), leading to a Fermi energy mismatch between the metal and semiconductor as explained in Fig. 9(a) and (b) (see supplementary data). This phenomenon causes a reduction in the band gap after sputtering Ag/Pt onto zirconia and enhances photoluminescence. Due to the wide band gap of zirconia, electrons require higher energy to transition from the valence band to the conduction band.

The energy and electron transfer mechanism is elucidated based on MSI involving $ZrO_2$ and Pt-Ag NPs. When the thin film atoms are excited by suitable electromagnetic spectrum wavelengths, electrons from the valence band gain sufficient energy to jump to the conduction band. During the emission cycle, electrons can reach deep trap states before returning to the ground state, with some emission energy exciting Ag-Pt atoms, which readily provide electrons to the zirconia conduction band. This process explains the observed reduction in band gap and enhancement in photoluminescence spectra at specific energy levels. The emissions primarily consist of band gap emissions and deep trap emissions in this context.

## 4. Discussions

During the sputtering process, the ratio of Zr to $O_2$ in the zirconia thin film varies, slightly influencing the behavior of Pt/Ag-Pt sputter deposition on the film. Optimal film quality is typically achieved with atomic % ratios of approximately $17.82 \pm 3\%$ for Zr and $78.53 \pm 3\%$ for $O_2$. The condensation mechanism of the sputtered compounds also significantly affects thin film quality.

Pt and Ag sputtered onto the zirconia thin film interact extensively with the crystal field, enhancing the film's emission characteristics. The addition of Ag to the surface enhances the properties of Pt NPs, making the ternary combination of Ag-Pt/$ZrO_2$ more optically active than Pt-$ZrO_2$ alone. Ag nanoparticles exhibit strong light interaction, with conduction electrons undergoing collective oscillations when excited by appropriate wavelengths. This enhancement primarily occurs through electron-phonon coupling and induced localized surface plasmon resonance (LSPR) of Ag-NPs.

Photoluminescence (PL) analysis of Pt/Ag-Pt sputtered zirconia thin films reveals new sharp peaks in the visible region of the spectra. The deposition rate of Pt/Ag-Pt over zirconia directly impacts film quality, with increased deposition time altering nanoparticle plasmonic properties and enhancing optical characteristics.

SEM analysis confirms that the nanoparticle size ranges are suitable for LSPR, with roughness and particle size playing crucial roles in tuning optical properties and enhancing photoluminescence in the thin film. For instance, in samples S2-S4, continuous nucleation and growth of platinum reduce surface roughness, while samples S5-S7 show decreased root-mean-square (RMS) roughness with increased Pt-Ag NP deposition time.

Overall, Pt/Ag-Pt sputtered zirconia thin films demonstrate enhanced optical properties due to these factors, making them promising for optoelectronic devices. Such coatings can increase device lifespan and efficiency, while remaining cost-effective for various applications in electronic diodes and optoelectronics.

## 6. Conclusion

Zirconia ($ZrO_2$) thin films, as well as Pt-$ZrO_2$ and Ag-Pt-$ZrO_2$ thin films, have been successfully deposited using both DC and RF power supplies at varying deposition times. The rate at which noble metals are deposited onto zirconia thin films plays a crucial role in determining film quality and enhancing their photoluminescence properties. These enhanced thin films hold promise for a wide range of applications.

One significant advantage of these thin films is their high heat-wearing capacity, attributed to good particle adhesion on various substrates. This characteristic makes them suitable for coating optoelectronic devices, where durability under thermal stress is essential. The optical properties of these plasmonic thin films, particularly their localized surface plasmon resonance (LSPR) capabilities, enable numerous applications in photovoltaic devices, optical components, and biosensors.

Pt/Pt-Ag embedded zirconia nanoparticles exhibit intriguing photoactive properties suitable for UV light emitters, blue and cyan light emitters, and other types of light-emitting diodes (LEDs). These materials can potentially increase the work function of devices, leading to higher conversion efficiency. Additionally, their cost-effective fabrication process makes them attractive for thin film coatings on solar panels and other optoelectronic devices, offering solutions to mitigate heat-related wear and tear issues.

In summary, Pt/Pt-Ag embedded zirconia thin films represent advanced materials with enhanced optical and photoactive properties, making significant strides in various technological applications due to their robustness and efficiency in optoelectronic devices.

**References**

[1] R.C. Garvie, R.H. Hannink, R.T. Pascoe, Ceramic Steel, Reprinted from Nature, 258(5537), 1975, pp.703-4.

[2] C. Piconi, G. Maccauro, Zirconia as a ceramic biomaterial, 1999.

[3] C. Yang, G. Zhang, N. Xu, J. Shi, Preparation and application in oil water separation of $ZrO_2$/$Al_2O_3$ MF membrane. ISBNNO: 03767388/98.

[4] J. Wang, W. Yin, X. He, Q. Wang, M. Guo, S. Chen, Good Biocompatibility and Sintering Properties of Zirconia Nanoparticles Synthesized via Vapor-phase Hydrolysis, Sci Rep 6 (2016). https://doi.org/10.1038/srep35020.

[5] J.A. Wang, M.A. Valenzuela, J. Salmones, A. Vázquez, A. García-Ruiz, X. Bokhimi, Comparative study of nanocrystalline zirconia prepared by precipitation and sol-gel methods, 2001.

[6] M. Varshney, A. Sharma, K.H. Chae, S. Kumar, S.O. Won, Electronic structure and dielectric properties of $ZrO_2$-$CeO_2$ mixed oxides, Journal of Physics and Chemistry of Solids 119 (2018).https://doi.org/10.1016/j.jpcs.2018.04.007.

[7] R. Mueller, R. Jossen, H.K. Kammler, S.E. Pratsinis, M.K. Akhtar, Growth of zirconia particles made by flame spray pyrolysis, AIChE Journal 50 (2004) 3085–3094. https://doi.org/10.1002/aic.10272.

[8] R.R. Piticescu, C. Monty, D. Taloi, A. Motoc, S. Axinte, Hydrothermal synthesis of zirconiananomaterials, n.d. www.elsevier.com/locate/jeurceramsoc.

[9] L. Liu, S. Wang, B. Zhang, G. Jiang, J. Yang, Supercritical hydrothermal synthesis of nano-$ZrO_2$:Influence of technological parameters and mechanism, J Alloys Compd 898 (2022). https://doi.org/10.1016/j.jallcom.2021.162878.

[10] S. Ben Amor, B. Rogier, G. Baud, M. Jacquet, M. Nardin, Characterization of zirconia filmsdeposited by r.f. magnetron sputtering, 1998.

[11] E. Çiftyürek, C.D. McMillen, K. Sabolsky, E.M. Sabolsky, Platinum-zirconium composite thin filmelectrodes for high-temperature micro-chemical sensor applications, Sens Actuators B Chem 207 (2015) 206–215. https://doi.org/10.1016/j.snb.2014.10.037.

[12] J.H. Shim, C.C. Chao, H. Huango, F.B. Prinz, Atomic layer deposition of yttria-stabilized zirconia for solid oxide fuel cells, Chemistry of Materials 19 (2007) 3850–3854. https://doi.org/10.1021/cm070913t.

[13] F. Yıldırım, S. Khalili, Z. Orhan, H.M. Chenari, Aydoğan, Highly sensitive self-powered UV-visiblephotodetector based on ZrO2-RGO nanofibers/n-Si heterojunction, J Alloys Compd 935 (2023). https://doi.org/10.1016/j.jallcom.2022.168054.

[14] B.E. Park, I.K. Oh, C. Mahata, C.W. Lee, D. Thompson, H.B.R. Lee, W.J. Maeng, H. Kim, Atomiclayer deposition of Y-stabilized ZrO2 for advanced DRAM capacitors, J Alloys Compd 722 (2017)307–312. https://doi.org/10.1016/j.jallcom.2017.06.036.

[15] M. Ismail, H. Abbas, C. Choi, S. Kim, Stabilized and RESET-voltage controlled multi-level switchingcharacteristics in ZrO2-based memristors by inserting a-ZTO interface layer, J Alloys Compd 835 (2020). https://doi.org/10.1016/j.jallcom.2020.155256.

[16] L. Gong, X. Wang, H. Hu, X. Xu, J. Zhao, Preparation and photoluminescence properties of ZrO2nanotube array-supported Eu3+doped ZrO2composite films, J Alloys Compd 705 (2017) 675–682. https://doi.org/10.1016/j.jallcom.2017.02.186.

[17] L. Renuka, K.S. Anantharaju, S.C. Sharma, H. Nagabhushana, Y.S. Vidya, H.P. Nagaswarupa, S.C. Prashantha, A comparative study on the structural, optical, electrochemical and photocatalytic properties of ZrO2nanooxide synthesized by different routes, J Alloys Compd 695 (2017) 382–395.https://doi.org/10.1016/j.jallcom.2016.10.126.

[18] R. Bacani, T.S. Martins, D.G. Lamas, M.C.A. Fantini, Structural studies of mesoporous ZrO_CeO and SiO mixed oxides for catalytical applications, (2015). https://doi.org/10.1016/j.jallcom.2016.01.213.

[19] W. Qi, J. Zhao, W. Zhang, Z. Liu, M. Xu, S. Anjum, S. Majeed, G. Xu, Visual and surface plasmon resonance sensor for zirconium based on zirconium-induced aggregation of adenosine triphosphate-stabilized gold nanoparticles, Anal Chim Acta 787 (2013) 126–131. https://doi.org/10.1016/j.aca.2013.05.030.

[20] Z. Cheng, N. Javed, D.M. Ocarroll, Optical and Electrical Properties of Organic Semiconductor ThinFilms on Aperiodic Plasmonic Metasurfaces, ACS Appl Mater Interfaces 12 (2020) 35579–35587. https://doi.org/10.1021/acsami.0c07099.

[21] S. Kunwar, M. Sui, P. Pandey, Z. Gu, S. Pandit, J. Lee, Improved Configuration and LSPR Responseof Platinum Nanoparticles via Enhanced Solid State Dewetting of In-Pt Bilayers, Sci Rep 9 (2019). https://doi.org/10.1038/s41598-018-37849-0.

[22] J. Bornacelli, H.G. Silva-Pereyra, L. Rodríguez-Fernández, M. Avalos-Borja, A. Oliver, From photoluminescence emissions to plasmonic properties in platinum nanoparticles embedded in silica by ion implantation, J Lumin 179 (2016) 8–15. https://doi.org/10.1016/j.jlumin.2016.06.032.

[23] Z. Cheng, N. Javed, D.M. Ocarroll, Optical and Electrical Properties of Organic Semiconductor ThinFilms on Aperiodic Plasmonic Metasurfaces, ACS Appl Mater Interfaces 12 (2020) 35579–35587. https://doi.org/10.1021/acsami.0c07099.

[24] C. Clavero, Plasmon-induced hot-electron generation at nanoparticle/metal-oxide interfaces forphotovoltaic and photocatalytic devices, Nat Photonics 8 (2014) 95–103. https://doi.org/10.1038/nphoton.2013.238.

[25] J. Bornacelli, H.G. Silva-Pereyra, L. Rodríguez-Fernández, M. Avalos-Borja, A. Oliver, From photoluminescence emissions to plasmonic properties in platinum nanoparticles embedded in silica by ion implantation, J Lumin 179 (2016) 8–15. https://doi.org/10.1016/j.jlumin.2016.06.032.

[26] R. Zhang, L. Bursi, J.D. Cox, Y. Cui, C.M. Krauter, A. Alabastri, A. Manjavacas, A. Calzolari, S. Corni, E. Molinari, E.A. Carter, F.J. García De Abajo, H. Zhang, P. Nordlander, How to Identify Plasmons from the Optical Response of Nanostructures, ACS Nano 11 (2017) 7321–7335. https://doi.org/10.1021/acsnano.7b03421.

[27] S. Kunwar, M. Sui, P. Pandey, Z. Gu, S. Pandit, J. Lee, Improved Configuration and LSPR Responseof Platinum Nanoparticles via Enhanced Solid State Dewetting of In-Pt Bilayers, Sci Rep 9 (2019). https://doi.org/10.1038/s41598-018-37849-0.

[28] M. Cueto, M. Piedrahita, C. Caro, B. Martínez-Haya, M. Sanz, M. Oujja, M. Castillejo, Platinum nanoparticles as photoactive substrates for mass spectrometry and spectroscopy sensors, Journal of Physical Chemistry C 118 (2014) 11432–11439. https://doi.org/10.1021/jp500190m.

[29] S. Veziroglu, M. Ullrich, M. Hussain, J. Drewes, J. Shondo, T. Strunskus, J. Adam, F. Faupel, O.C. Aktas, Plasmonic and non-plasmonic contributions on photocatalytic activity of Au-TiO2 thin film under mixed UV–visible light, Surf Coat Technol 389 (2020). https://doi.org/10.1016/j.surfcoat.2020.125613.

[30] C.C. Lee, C.C. Kuo, Optical coatings for displays and lighting, in: Optical Thin Films and Coatings: From Materials to Applications, Elsevier Ltd, 2013: pp. 564–595. https://doi.org/10.1533/9780857097316.4.564.

[31]  G. Muralidharan, D.F. Wilson, M.L. Santella, D.E. Holcomb, Cladding Alloys for Fluoride Salt Compatibility Final Report, Oak Ridge, TN (United States), 2011. https://doi.org/10.2172/1006466.

[32]  N. Nafarizal, Precise Control of Metal Oxide Thin Films Deposition in Magnetron Sputtering Plasmasfor High Performance Sensing Devices Fabrication, Procedia Chem 20 (2016) 93–97. https://doi.org/10.1016/j.proche.2016.07.016.

[33]  N. Shimosako, H. Sakama, Basic photocatalytic activity of ZrO2 thin films fabricated by a sol-gel method under UV-C irradiation, Thin Solid Films 732 (2021). https://doi.org/10.1016/j.tsf.2021.138786.

[34]  N.C. Horti, M.D. Kamatagi, S.K. Nataraj, M.S. Sannaikar, S.R. Inamdar, Photoluminescence properties of zirconium oxide (ZrO2) nanoparticles, in: AIP Conf Proc, American Institute of PhysicsInc., 2020. https://doi.org/10.1063/5.0022460.

[35]  E. Çiftyürek, C.D. McMillen, K. Sabolsky, E.M. Sabolsky, Platinum-zirconium composite thin filmelectrodes for high-temperature micro-chemical sensor applications, Sens Actuators B Chem 207 (2015) 206–215. https://doi.org/10.1016/j.snb.2014.10.037.

[36]  N.C. Horti, M.D. Kamatagi, S.K. Nataraj, M.S. Sannaikar, S.R. Inamdar, Photoluminescence properties of zirconium oxide (ZrO2) nanoparticles, in: AIP Conf Proc, American Institute of PhysicsInc., 2020. https://doi.org/10.1063/5.0022460.

[37]  M. Varshney, A. Sharma, K.H. Chae, S. Kumar, S.O. Won, Electronic structure and dielectric properties of ZrO2-CeO2 mixed oxides, Journal of Physics and Chemistry of Solids 119 (2018) 242–250. https://doi.org/10.1016/j.jpcs.2018.04.007.

[38]  S.K. Rawal, R. Chandra, Wettability and Optical Studies of Films Prepared from Power Variation ofCo-sputtered Cr and Zr Targets by Sputtering, Procedia Technology 14 (2014) 304–311. https://doi.org/10.1016/j.protcy.2014.08.040.

[39]  I. Farid, A. Boruah, J. Chutia, A.R. Pal, H. Bailung, Low loaded platinum (Pt) based binary catalystselectrode for PEMFC by plasma co-sputtered deposition method, Mater Chem Phys 236 (2019). https://doi.org/10.1016/j.matchemphys.2019.121796.

[40]  G. V. Hartland, Optical studies of dynamics in noble metal nanostructures, Chem Rev 111 (2011)3858–3887. https://doi.org/10.1021/cr1002547.

[41]  N. Soin, S.S. Roy, L. Karlsson, J.A. McLaughlin, Sputter deposition of highly dispersed platinum nanoparticles on carbon nanotube arrays for fuel cell electrode material, Diam Relat Mater 19 (2010)595–598. https://doi.org/10.1016/j.diamond.2009.10.029.

[42] J. Tauc, Optical properties and electronic structure of amorphous Ge and Si, Mater Res Bull 3 (1968) 37–46. https://doi.org/10.1016/0025-5408(68)90023-8.

[43] Q. Zhu, J. Lu, Y. Wang, F. Qin, Z. Shi, C. Xu, Burstein-Moss Effect Behind Au Surface Plasmon Enhanced Intrinsic Emission of ZnO Microdisks, Sci Rep 6 (2016) 36194. https://doi.org/10.1038/srep36194.

[44] Q. Zhu, J. Lu, Y. Wang, F. Qin, Z. Shi, C. Xu, Burstein-Moss Effect Behind Au Surface Plasmon Enhanced Intrinsic Emission of ZnO Microdisks, Sci Rep 6 (2016). https://doi.org/10.1038/srep36194.

[45] D. Manoharan, A. Loganathan, V. Kurapati, V.J. Nesamony, Unique sharp photoluminescence of size-controlled Sono chemically synthesized zirconia nanoparticles, Ultrason Sonochem 23 (2015) 174–184. https://doi.org/10.1016/j.ultsonch.2014.10.004.

[46] X.H. Wang, Z.C. Su, J.Q. Ning, M.Z. Wang, S.J. Xu, S. Han, F. Jia, D.L. Zhu, Y.M. Lu, Influence of lattice vibrations on luminescence and transfer of excitons in WS2 monolayer semiconductors, J Phys D Appl Phys 49 (2016). https://doi.org/10.1088/0022-3727/49/46/465101.

[47] M. Singh Gangwar, P. Agarwal, Plasmon-enhanced photoluminescence and Raman spectroscopy of silver nanoparticles grown by solid state dewetting IOP Publishing Journal (2023). https://doi.org/10.1088/1361.

[48] M. Kumar, G.B. Reddy, Tailoring surface plasmon resonance in Ag: ZrO2 nanocomposite thin films, Physica E Low Dimens Syst Nanostructure 43 (2010) 470–474. https://doi.org/10.1016/j.physe.2010.08.031.

[49] L. Skowronski, R. Chodun, M. Trzcinski, K. Zdunek, Optical Properties of Amorphous Carbon Thin Films Fabricated Using a High-Energy-Impulse Magnetron-Sputtering Technique, Materials 16 (2023) 7049. https://doi.org/10.3390/ma16217049.

[50] T. Guo, M.N. Karim, K. Ghosh, M.M. Murshed, K. Rezwan, M. Maas, Plasmonic porous ceramics based on zirconia-toughened alumina functionalized with silver nanoparticles for surface-enhanced Raman scattering, Open Ceramics 9 (2022). https://doi.org/10.1016/j.oceram.2022.100228.

[51] Z. Cheng, N. Javed, D.M. Ocarroll, Optical and Electrical Properties of Organic Semiconductor Thin Films on Aperiodic Plasmonic Meta surfaces, ACS Appl Mater Interfaces 12 (2020) 35579–35587. https://doi.org/10.1021/acsami.0c07099.

[52] M. Mamivand, M.A. Zaeem, H. El Kadiri, L.Q. Chen, Phase field modeling of the tetragonal-to-monoclinic phase transformation in zirconia, Acta Mater 61 (2013) 5223–5235. https://doi.org/10.1016/j.actamat.2013.05.015.

[53] Y. Li, Y. Zhang, K. Qian, W. Huang, Metal-Support Interactions in Metal/Oxide Catalysts and Oxide-Metal Interactions in Oxide/Metal Inverse Catalysts, ACS Catal 12 (2022) 1268–1287. https://doi.org/10.1021/acscatal.1c04854.